\documentclass[preprint,12pt]{elsarticle}

\usepackage{graphicx}
\usepackage{amssymb}

\usepackage{lineno}
\usepackage{soul}
\usepackage{color}
\usepackage{multirow}
\usepackage{rotating}
\usepackage{xcolor}
\usepackage{xargs}                      

\usepackage[colorinlistoftodos,prependcaption,textsize=tiny]{todonotes}

\usepackage{url}

\usepackage[colorlinks=true,linkcolor=blue,citecolor=blue]{hyperref}

\journal{Medical Image Analysis}

\begin{document}

\begin{frontmatter}


\title{Neuropsychiatric Disease Classification Using Functional Connectomics - Results of the Connectomics in NeuroImaging Transfer Learning Challenge}



\author[1,2,ukb]{Markus D. Schirmer\corref{cor1}\fnref{cni}}
\ead{markus.schirmer@ukbonn.de}
\cortext[cor1]{Corresponding authors} 
\fntext[fn1]{These authors are the organizers and contributors to the setup of the Challenge.}

\author[3]{Archana Venkataraman\fnref{cni}}
\author[4,10]{Islem Rekik\fnref{cni}}
\author[5]{Minjeong Kim\fnref{cni}}
\author[7,8,9]{Stewart H. Mostofsky}
\author[7,8]{Mary Beth Nebel}
\author[7,9,12]{Keri Rosch}
\author[7,9]{Karen Seymour}
\author[7]{Deana Crocetti}
\author[S1-1,S1-2]{Hassna Irzan}
\author[S1-2]{Michael H\"utel}
\author[S1-2]{Sebastien Ourselin}
\author[S1-3]{Neil Marlow}
\author[S1-2,S1-1]{Andrew Melbourne}
\author[S2]{Egor Levchenko}
\author[DCSUoS]{Shuo Zhou}
\author[DCSUoS]{Mwiza Kunda}
\author[DCSUoS]{Haiping Lu}
\author[S5-1,S5-2]{Nicha C. Dvornek}
\author[S5-2]{Juntang Zhuang}
\author[6]{Gideon Pinto\fnref{cni}}
\author[6]{Sandip Samal\fnref{cni}}
\author[6]{Jennings Zhang\fnref{cni}}
\author[11]{Jorge L. Bernal-Rusiel\fnref{cni}}
\author[6,12]{Rudolph Pienaar\fnref{cni}}
\author[6,13]{Ai Wern Chung\corref{cor1}\fnref{cni}}
\ead{aiwern.chung@childrens.harvard.edu}

\address[1]{Massachusetts General Hospital, Harvard Medical School, Boston, USA}
\address[2]{German Center for Neurodegenerative Diseases, Bonn, Germany}
\address[ukb]{Clinic for Neuroradiology, University Hospital Bonn, Germany}
\address[3]{Department of Electrical and Computer Engineering, Johns Hopkins University, Baltimore, USA}
\address[4]{BASIRA lab, Faculty of Computer and Informatics, Istanbul Technical University, Istanbul, Turkey}
\address[10]{School of Science and Engineering, Computing, University of Dundee, UK}
\address[5]{Department of Computer Science, University of North Carolina at Greensboro, USA}
\address[7]{Center for Neurodevelopmental and Imaging Research, Kennedy Krieger Institute, Baltimore, USA}
\address[8]{Department of Neurology, Johns Hopkins School of Medicine, USA}
\address[9]{Department of Psychiatry and Behavioral Sciences, Johns Hopkins School of Medicine, Baltimore, USA}
\address[12]{Department of Neuropsychology, Kennedy Krieger Institute, Baltimore, USA}
\address[S1-1]{Department of Medical Physics and Biomedical Engineering, University College London, UK}
\address[S1-2]{School of Biomedical Engineering and Imaging Sciences, King’s College London, UK}
\address[S1-3]{Institute for Women’s Health, University College London, UK}
\address[S2]{Institute for Cognitive Neuroscience, Higher School of Economics, Moscow, Russia}
\address[DCSUoS]{Department	of Computer Science, The University of Sheffield, Sheffield, UK}
\address[S5-1]{Department of Radiology \& Biomedical Imaging, Yale University, New Haven, CT, USA}
\address[S5-2]{Department of Biomedical Engineering, Yale University, New Haven, CT, USA}
\address[6]{Fetal-Neonatal Neuroimaging and Developmental Science Center, Division of Newborn Medicine, Boston Children's Hospital, Harvard Medical School, Boston, MA, USA}
\address[11]{Teradyte LLC, Coral Gables, FL, USA}
\address[12]{Department of Radiology, Boston Children's Hospital,Harvard Medical School, Boston, MA, USA}
\address[13]{Department of Pediatrics, Boston Children's Hospital, Harvard Medical School, Boston, MA, USA}

\begin{abstract}
Large, open-source datasets, such as the Human Connectome Project and the Autism Brain Imaging Data Exchange, have spurred the development of new and increasingly powerful machine learning approaches for brain connectomics. However, one key question remains: are we capturing biologically relevant and generalizable information about the brain, or are we simply overfitting to the data? To answer this, we organized a scientific challenge, the Connectomics in NeuroImaging Transfer Learning Challenge (CNI-TLC), held in conjunction with MICCAI 2019. CNI-TLC included two classification tasks: (1) diagnosis of Attention-Deficit/Hyperactivity Disorder (ADHD) within a pre-adolescent cohort; and (2) transference of the ADHD model to a related cohort of Autism Spectrum Disorder (ASD) patients with an ADHD comorbidity. In total, 240 resting-state fMRI (rsfMRI) time series averaged according to three standard parcellation atlases, along with clinical diagnosis, were released for training and validation (120 neurotypical controls and 120 ADHD). We also provided Challenge participants with demographic information of age, sex, IQ, and handedness. The second set of 100 subjects (50 neurotypical controls, 25 ADHD, and 25 ASD with ADHD comorbidity) was used for testing. Classification methodologies were submitted in a standardized format as containerized Docker images through ChRIS, an open-source image analysis platform. Utilizing an inclusive approach, we ranked the methods based on 16 metrics: accuracy, area under the curve, F1-score, false discovery rate, false negative rate, false omission rate, false positive rate, geometric mean, informedness, markedness, Matthew’s correlation coefficient, negative predictive value, optimized precision, precision, sensitivity, and specificity. The final rank was calculated using the rank product for each participant across all measures. Furthermore, we assessed the calibration curves of each methodology. Five participants submitted their method for evaluation, with one outperforming all other methods in both ADHD and ASD classification. However, further improvements are still needed to reach the clinical translation of functional connectomics. We have kept the CNI-TLC open as a publicly available resource for developing and validating new classification methodologies in the field of connectomics.
\end{abstract}


\end{frontmatter}


\section{Introduction}
\label{S:1}

Functional connectomics, or the study of whole-brain synchronization maps, has become of increasing interest to the neuroscientific community in recent years. For example, functional connectomics has provided valuable insight into human cognition (\cite{MILL2017124, RN126}), the system-level organization of the brain over development and aging (\cite{SOMERVILLE2018456, RN125}), and whole-brain functional alterations in disease or injury (\cite{DSouza3,DSouza_NI,VenkIEEE12,VenkIEEE13,VenkIEEE16,VenkNICL,ktena2019brain,Bonkhoff2020}). Large, open-source initiatives, such as the Human Connectome Project (\cite{VanEssen}) and the Autism Brain Imaging Data Exchange (\cite{di2014autism}), have spurred the development of new and increasingly powerful machine learning strategies to capitalize on these resources. 

One popular goal in connectomics is to classify patients from controls. However, objective comparisons of these algorithms across studies can be challenging, due to variations in image acquisition, preprocessing pipeline, and the specific cohort under consideration (\cite{abraham2017deriving,pervaiz2019optimising}). Given these factors, the question of whether a proposed model is capturing biologically relevant and generalizable information about the brain, or simply overfitting to the data, remains to be investigated. In addition to data inconsistencies, performance is assessed using a restricted and non-standardized subset of evaluation metrics, further hindering comparisons across studies (\cite{maier-hein2018rankcomp}). Similar to other fields, scientific challenges provide a way to control for these issues and have been conducted in various domains, such as image registration (\cite{murphy2011evaluation}), lesion segmentation (\cite{heimann2009comparison,menze2014multimodal,mendrik2015mrbrains,commowick2018objective,CARASS201777}), and estimation of clinical scores (\cite{wolterink2016evaluation}). 

The Connectomics in NeuroImaging Transfer Learning Challenge (CNI-TLC) described here tackles the issues of generalizability and clinical relevance of functional connectomes by leveraging unique resting-state functional MRI (rsfMRI) datasets of Attention-Deficit/Hyperactivity Disorder (ADHD), Autism Spectrum Disorder (ASD), and Neurotypical Controls (NC). ADHD is a chronic neurobehavioral disorder characterized by inattention, hyperactivity, and impulsivity that affects more than 6 million children worldwide (\cite{hamed2015diagnosis,DSM5}). In contrast, ASD patients typically exhibit problems with social skills, communication, and abnormal behavioral habits (\cite{huerta2012diagnostic,Pelphrey,Dowell,McPart2,DSM5}). While the hallmark behavioral manifestation of ADHD and ASD cohorts differ dramatically, there is a significant comorbidity between the two disorders (\cite{leitner2014co}). Inspired by this finding, we have developed a challenge framework to investigate whether connetome-based features identified in ADHD patients can be transferred to an ASD population for a classification task. Our Challenge setup thereby extends the conventional notion of ``transfer learning''. As opposed to transferring a previously optimized model for re-training and testing on a new cohort,  we propose to transfer and test the learned representation itself, without additional training. This allows us to test whether a robust symptomatology can be learned for ADHD, and if so, whether it can also be extracted in a co-morbid population. Participants were asked to design an ADHD versus NC classification model using rsfMRI time series and demographic measures from 100 and 20 examples of each class for training and validation, respectively. In Task I of the Challenge, we evaluated the classifiers on withheld ADHD and NC data (25 examples of each class). In Task II of the Challenge, we assessed the classification performance of the ADHD model on ASD patients, who have been diagnosed with an ADHD comorbidity (25 examples of each class). Our unique Challenge assesses both the ability of the method to extract functional connectivity patterns related to ADHD symptomatology and how much of this information “transfers” across clinical domains with an overlapping diagnosis. Evaluation was performed using 16 different metrics, while applying cross-validation on the test data, and participants were ranked relative to one another. In this paper, we describe the Challenge data, organization, and detailed evaluation. We also describe the methodology submitted by each participant and the corresponding experimental results.

\section{Materials and Methods}
\subsection{Patient Population}
The data used for CNI-TLC were amassed retrospectively across multiple studies conducted by the Center for Neurodevelopmental and Imaging and Research (CNIR) at the Kennedy Krieger Institute (KKI) in Baltimore, MD (see~\ref{app:A} for the list of study names and Johns Hopkins IRB approval numbers). The overall cohort includes 145 children diagnosed with ADHD, 25 children with a primary diagnosis of ASD who also meet the diagnostic criteria for ADHD, and 170 NC. All children are between 8-12 years of age and are considered high-functioning based on having a full-scale IQ at or above the normal range; the groups have been matched on age and full-scale IQ. Detailed cohort characteristics are summarized in Table~\ref{tab:cohort}.

\bgroup
\linespread{1.2}
\begin{table}[ht!]
    \centering
    \caption{Cohort characterization. For each sample (Training, Validation, and Testing), Controls were matched to patients for all demographics available ($p>0.05$). FSIQ: Wechsler Intelligence Scale for Children Full Scale Intelligence Quotient; EH: Edinburgh Handedness. SD: Standard Deviation} 
    \label{tab:cohort}
    \tiny
    \begin{tabular}{|c|c|c|c|c|c|c|c|c|c|}
    \hline
         & \multirow{2}{*}{\textbf{All}} & \multicolumn{2}{c|}{\textbf{Training}}  & \multicolumn{2}{c|}{\textbf{Validation}}  & \multicolumn{4}{c|}{\textbf{Testing}}\\ \cline{3-10}
         & & \multirow{2}{*}{\textbf{ADHD}} & \multirow{2}{*}{\textbf{Control}} & \multirow{2}{*}{\textbf{ADHD}} & \multirow{2}{*}{\textbf{Control}} & \multicolumn{2}{c|}{\textbf{Task I) ADHD}} & \multicolumn{2}{c|}{\textbf{Task II) ASD}} \\ \cline{7-10}
         &&&&&& \textbf{Patients} & \textbf{Control} & \textbf{Patients} & \textbf{Control}  \\ \hline
         n & 340 & 100 &100 &20 &20 &25&25&25&25 \\ \hline
         Sex & 239  & 70 & 69 & 14 & 14 & 18 & 16 & 17 & 21 \\
         (Male; \%) & (70.3) &  (70.0) &  (69.0) &  (70.0) &  (70.0) &  (72.0) &  (64.0) &  (68.0) &  (84.0) \\ \hline
         Age (years) & 10.4 & 10.4 & 10.3 & 10.3 & 10.2 & 10.3 & 10.8 & 10.6 & 10.7 \\
         (mean (SD)) & (1.3) & (1.5) & (1.2) & (1.4) & (1.2) & (1.3) & (0.9) & (1.4) & (1.4) \\ \hline 
         FSIQ & 111.3 & 109.2 & 115.4 & 104.8 & 113.7 & 105.4 & 111.0 & 116.2 & 108.0 \\
         (mean (SD)) & (12.7) & (12.2) & (10.4) & (13.4) & (14.9) & (12.3) & (11.0) & (14.1) & (14.9) \\ \hline
         EH & 0.7 & 0.7 & 0.7 & 0.7 & 0.7 & 0.7 & 0.8 & 0.6 & 0.7 \\ 
         (mean (SD)) & (0.5) & (0.5) & (0.5) & (0.5) & (0.4)  & (0.6) & (0.3) & (0.7) & (0.6) \\ \hline
    \end{tabular}
\end{table}
\egroup

\subsection{Clinical Assessment}

Participants received an ADHD diagnosis if they met criteria for ADHD using either the Diagnostic Interview for Children and Adolescents (DICA), Fourth Edition (\cite{reich1997}) or the Kiddie Schedule for Affective Disorders and Schizophrenia (K-SADS) for School-Aged Children-Present and Lifetime Version (\cite{Kaufman}), in addition to either (1) a t-score of 60 on the Inattentive or Hyperactive subscales of the Conners' Parent or Teacher Rating Scales-Revised Long Version or the Conners-3 (\cite{connors2002,connors2008}), or (2) a score of 2 on at least 6 items on the Inattentive or Hyperactivity/Impulsivity scales of the ADHD Rating Scale-IV, Home or School Versions (\cite{dupaul1998}). These tests are designed for children at the age of 6-18 years and were administered by a trained psychologist in CNIR. No additional instructions were given to either the children or examiners for the purposes of this challenge. 

Diagnostic criteria for ASD was assessed via two standard instruments: the Autism Diagnostic Observation Schedule Version 2 (ADOS-2) (\cite{Gotham}) and the Autism Diagnostic Interview-Revised (ADI-R) (\cite{Lord,Lord2}). Similar to K-SADS, the ADOS-2 evaluation consists of both structured questions and unstructured narratives. ADOS-2 is used to quantify both socio-communication deficits, as well as repeated/repetitive behaviors. Once again, ADOS-2 was administered by a trained psychologist in CNIR. In contrast, ADI-R is a written questionnaire for parents and/or caregivers about the child. ADI-R provides categorical results for three domains: language and communication, reciprocal social interactions, and repetitive behaviors. The overall ASD diagnosis is based on the summed dimensional scores for each battery. 

Inclusion in the NC group required not meeting criteria for any diagnosis on the DICA or K-SADS, having scores below clinical cut-offs on the parent and teacher (when available) Conners' and ADHD Rating Scales, as well as having no immediate family members diagnosed with ADHD or ASD.

In addition to diagnosis, we provided four demographic variables to challenge participants. These variables are age, sex, the Wechsler Intelligence Scale for Children (fourth or fifth editions) Full Scale Intelligence Quotient (FSIQ; \cite{WISC-IV}), and the Edinburgh Handedness index (\cite{Oldfield}).

\subsection{Data Acquisition and Preprocessing}
The rsfMRI data used in this challenge was acquired on a Philips 3T Achieva scanner housed in the F.M. Kirby Research Center for Functional Brain Imaging at KKI\footnote{http://www.kennedykrieger.org/kirby-research-center}. The acquisition protocol used a single shot, partially parallel gradient-recalled EPI sequence with TR/TE 2500/30ms, flip angle 70$^{\circ}$, and voxel resolution $3.05 \times 3.15 \times 3$mm$^3$. The scan duration was either 128 or 156 time samples. Children were instructed to relax with their eyes open and focus on a central cross-hair, while remaining still for the duration of the scan. All participants completed a mock scanning session to habituate to the MRI environment.

The rsfMRI data was preprocessed using an in-house pipeline developed by CNIR and implemented in SPM-12 (\cite{SPM}). The pipeline included slice timing correction, rigid body realignment, and normalization to the EPI version of the MNI template. The time courses were temporally detrended in order to remove gradual trends in the data. CompCorr (\cite{Behzadi}) was performed to estimate and remove spatially coherent noise from the white matter and ventricles, along with the linearly detrended versions of the six rigid body realignment parameters and their first derivatives. From here, the data was spatially smoothed with a 6mm FWHM Gaussian kernel and bandpass filtered between 0.01-0.1 Hz. Finally, spike correction was performed using the AFNI package (\cite{Cox}), as an alternative to motion scrubbing. 

\begin{figure}[bt!]
    \centering
    \includegraphics[width=\linewidth]{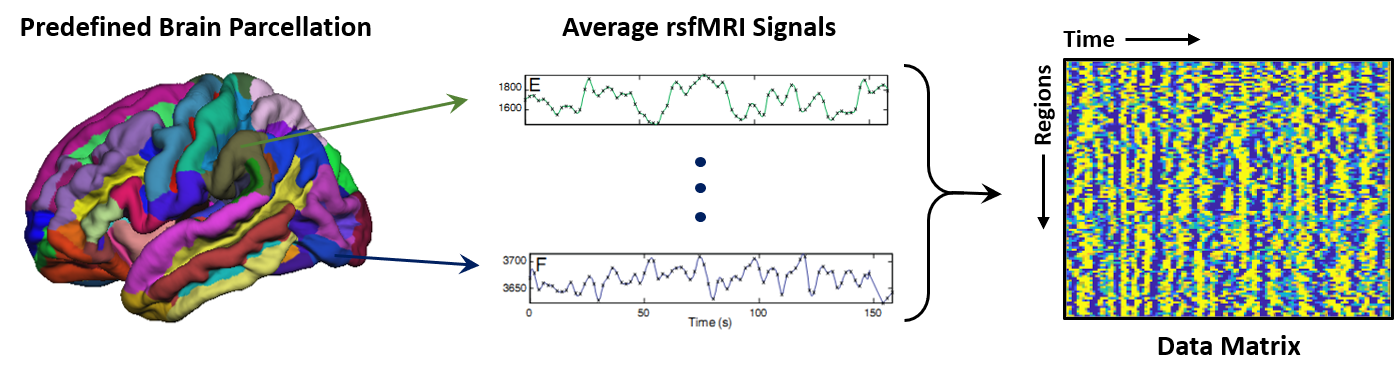}
    \caption{Pipeline of extracting data from a brain parcellation, which was provided for challenge participants.}
    \label{fig:pipeline}
\end{figure}

We released average region-wise time series data for the CNI-TL Challenge, from which participants could compute both static and dynamic connectivity measures. These average signals were computed based on three standard parcellations: (1) the AAL atlas (\cite{tzourio2002automated}), which consists of 90 cortical/subcortical regions and 26 cerebellar regions, (2) the Harvard-Oxford atlas (\cite{desikan2006automated}), which consists of 110 cerebral and cerebellar regions, and (3) the Craddock 200 atlas (\cite{craddock2012whole}), which is a finer parcellation of 200 regions. The choice of atlases enabled participants to analyze the rsfMRI data at multiple spatial scales. Fig.~\ref{fig:pipeline} illustrates our postprocessing workflow for a single parcellation. The mean time courses for each parcellation were aggregated into a single data file for participants to use at their discretion.

\subsection{Challenge Design}
\subsubsection{Data}
Submissions to CNI-TLC were evaluated based on two classification tasks: {\it Task I} - Primary Classification of ADHD versus NC; and {\it Task II} - Transference Classification of ASD versus NC. The cohort was divided into training (100 ADHD, 100 NC), validation (20 ADHD, 20 NC) and testing datasets (Task~I: 25 ADHD, 25 NC; Task~II: 25 ASD, 25 NC). The training and validation datasets only consisted of ADHD and NC subjects, in fulfillment of Task~I, and was made available for participants to download via the challenge website\footnote{\url{http://www.brainconnectivity.net}}. Training data was released in June 2019, and validation data followed thirty days later. Each dataset was organized into top directories, one for each subject. The subject directories included four .csv files, one containing the demographic variables, and a separate file for each parcellation. Testing datasets for both Task~I and Task~II (composed of NC, ADHD and ASD patients with ADHD comorbidity) were not released to the public. The patient and NC cohorts were matched on age, sex, FSIQ, and handedness in each of the three datasets.

\subsubsection{Submission Infrastructure}
Participants were instructed to submit their trained models as Docker images\footnote{\url{http://www.docker.com}}. Docker is an emerging Platform-As-A-Service (PaaS) product that provides a simple mechanism for bundling applications with all their required dependencies in easily isolated components called containers. In this manner, Docker containerized applications can be executed on all the major computing platforms (Linux, Mac, and Windows) with no additional software requirements besides Docker itself. By using Docker containers as the deployment vector for the trained models, the problem of actually running these models on a single evaluation system was addressed. 

For our Challenge, participants were directed to clone a GitHub repository\footnote{\url{https://github.com/aichung/pl-cni_challenge}}. This repository contained all the basic skeleton components for building a Docker image solution. In addition to the Docker components, a dummy stub Python program, \texttt{pl-cni\_challenge.py} was provided as a launching point for a solution. Participants could either code directly into \texttt{pl-cni\_challenge.py}, import their existing code as a Python module, or include a suitably compiled executable that can be called using the \texttt{os.system()} function in Python. 

The repository itself was created by running a cookie-cutter template code\footnote{\url{https://github.com/FNNDSC/cookiecutter-chrisapp}} that creates plug-ins for ChRIS\footnote{\url{https://chrisproject.org}}. The use of the ChRIS system had no significant impact on participants other than defining a standard command line interface contract. 

All submissions were required to accept two positional arguments -- an \texttt{inputDirectory} that would contain the data, and an \texttt{outputDirectory} that would store their model predictions. In addition, submitted solutions were to expect the input directory to contain test data with the same folder structure as the released training and validation data, with subject diagnosis information excluded. Participants were instructed to write two text files into the output directory: \texttt{classification.txt} containing binary classification labels for each subject, and \texttt{score.txt} containing the probability score of each corresponding label in \texttt{classification.txt}. Full instructions on how to create, compile and execute a compatible Docker image for submission were provided on the Challenge website. The Docker images were executed on each test subject, one at a time. For consistency, all submissions were evaluated on the same machine.

\subsection{Participants}
Five solutions were submitted before the deadline of the Challenge. A brief summary of each method is given below. Additional details for each method can be found in~\ref{app:B}.

\textbf{Submission 1 (S1).} \textit{MeInternational}. The method is based on building classification pipelines by using permutations of functional connectivity matrices,
anatomical atlases, and classification algorithms (\cite{abraham2017deriving}). From each atlas and time series, correlation, covariance, partial correlation, precision, and tangent embedding for functional connectivity were estimated. Classification was based on support vector machines (SVMs), linear regression ($l_1$ or $l_2$ regularization), random forest, $k$-nearest neighbor, and naive Bayes classifiers. Each combination was tested on the training data and evaluated on the validation data set, where the best performing pipeline was chosen based on its prediction accuracy score.

\textbf{Submission 2 (S2).} \textit{HSE}. The core principle of this submission was the utilization of eigenvalues of the normalized Laplacian. In brief, for each connectome, the normalized Laplacian and its corresponding eigenvalues were calculated. The full set of eigenvalues were subsequently used as features in an SVM classification algorithm using a polynomial kernel.

\textbf{Submission 3 (S3).} \textit{ShefML}. This solution consists of two stages: First, pairwise region of interest (ROI) features, ROI-to-ROI, were extracted from the rsfMRI time series by computing Tangent Pearson connectivity (\cite{kunda2020improving}). Then, an SVM was trained which is regularized by the statistical independence (\cite{gretton2005measuring}) between the classifier decision scores and three types of demographic information: gender, age, and handedness score (\cite{zhou2020side}).

\textbf{Submission 4 (S4).} \textit{ShefML}. In this approach, five types of features were extracted from the time series by computing: mean and standard deviation, Pearson correlation, Tangent (\cite{varoquaux2010detection}), covariance (\cite{varoquaux2010brain}), and Tangent Pearson (\cite{kunda2020improving}) connectivity. Subsequently, five classifiers (\cite{zhou2020side}), one per feature, were trained and classifications were combined by majority voting.

\textbf{Submission 5 (S5).} \textit{YaleIPAG}. Here, an LSTM-based (long short-term memory) network was used to learn directly from time-series data (\cite{Dvornek2017}). Twenty-two AAL ROIs were first selected based on consistent connectivity differences between ADHD and controls in bootstrapped samples. The time-series from these ROIs were input to an LSTM, with the demographic data used in hidden and cell state initialization (\cite{Dvornek2018a}).

\subsection{Evaluation and Ranking}
\subsubsection{Challenge Evaluation Metrics} 
Taking an inclusive approach, we assessed multiple measures commonly used in classification tasks, such as accuracy and area under the curve (AUC), along with distributional measures, such as geometric-mean and optimized precision. This approach provides an intuitive and robust characterization of each submission. A full list of the utilized measures and their interpretation is given in Table~\ref{tab:measures}, while a detailed description of these measures can be found elsewhere in the literature (\cite{hossin2015review}).

\bgroup
\linespread{1.2}
\begin{table}[ht!]
    \centering
    \caption{Summary of evaluation metrics, including abbreviation (Abbr.) used, a short description, and condition under which one participant outperforms another (‘Better if’).}
    \label{tab:measures}
    \tiny
    \begin{tabular}{|c|c|l|c|}
         \hline
        {\bf Measure} & {\bf Abbr.} & {\bf Description} & {\bf Better if} \\ \hline
        Accuracy & Acc & Ratio of correct predictions over the total number of & higher   \\ 
                 && instances evaluated & \\ \hline
        Area under curve & AUC & Reflects overall ranking performance of classifier & higher  \\ \hline
        F1-score & F1 & Harmonic mean between recall (sensitivity) and precision & higher \\ \hline
        False discovery rate & FDR & Fraction of misclassified positive samples in relation to the & lower \\ 
                             && number of the total positive classified samples & \\ \hline
        False negative rate & FNR & Also known as miss rate. Fraction of misclassified negative & lower \\
                            && samples in relation to the number of positive samples & \\ \hline
        False omission rate & FOR & Fraction of misclassified negative samples in relation to the & lower \\
                            && number of the total negative classified samples & \\ \hline
        False positive rate & FPR & Fraction of misclassified positive samples in relation to the & lower \\
                            && number of total negative classified samples & \\ \hline
        Geometric mean & GM & Geometric mean of sensitivity and precision & higher \\ \hline
        Informedness & Inf & Also known as Youden's J statistic. It summarizes the true & higher \\ 
                     && positive and true negative rates of a classifier & \\ \hline
        Markedness & Mark & Summarizes the positive (prediction) and negative predictive & higher \\
                   && value of a classifier & \\ \hline
        Matthews & MCC & Correlation coefficient between the observed and predicted & higher \\
        correlation && binary classification with values between -1 (total & \\ 
        coefficient && disagreement) and 1 (perfect prediction) & \\ \hline
        Negative predictive & NPV & Fraction of correctly classified negative samples in relation & higher \\
        value && to the number of the total negative classified samples & \\ \hline
        Optimized precision & OP & Measure aiming to simultaneously minimize the difference & higher \\
                            && in sensitivity and specificity, while maximizing their sum. & \\
                            && Precision is subsequently "corrected" by the ratio of this & \\
                            && difference and sum. & \\ \hline
        Precision & Pre & Fraction of correctly classified positive samples in relation to & higher \\
                  && the number of the total positive classified samples & \\ \hline
        Sensitivity & Sen & Fraction of positive samples that are correctly classified & higher \\ \hline
        Specificity & Spec & Fraction of negative samples that are correctly classified & higher \\ \hline
    \end{tabular}
\end{table}
\egroup

\subsubsection{Comparison and Ranking Strategy}
We first evaluated the model performances on the validation set to assess the primary classification task of ADHD versus NC. This evaluation provides a baseline measure of performance for each algorithm on data that has been made available to the participants. It also allows us to investigate the drop in performance when presenting the classifier with unseen test data later on.

We utilized a 5-fold cross-validation scheme for statistical testing of each submission on the unseen data. Namely, we ``randomly" divided our testing dataset into five equal sized, statistically indistinguishable (in terms of sex, age, FSIQ, and handedness), disjointed folds\footnote{https://github.com/mdschirmer/MDS;\cite{schirmer2019richclub}} and calculated the evaluation metrics (see Table~\ref{tab:measures}), using data from four of the five folds. This process yields five quantitative values for each evaluation metric. The cross validation procedure was repeated 100 times (on different random splits of the data), resulting in a distribution of 500 values for each metric. 

Our initial ranking was based on the median of the distributions, where the ranking between participants was statistically evaluated using a pairwise Wilcoxon test. Considering the total of 320 tests (ten comparisons between challengers with 16 measures each for two classification tasks, i.e., ADHD and ASD), we set the significance level to $p \leq 0.0001$ (Bonferroni correction of 0.05/320). Finally, we calculated the rank product (geometric mean) for each participant, which gave us the final ranking of submissions. 

In addition, we evaluated submissions based on calibration curves. With calibration curves, we can gain an idea of the model's behavior and confidence in performing the classification tasks. These curves assess if the model can reliably estimate the probability of the diagnosis by plotting the mean predicted probability against the true probability for each user-specified probability bin. Here, we summarize the results of the cross-validation approach as a single calibration curve using 10 bins of width 0.1, and fit a linear model to the predicted probability against the observed probability for comparison. 

\section{Results}
\label{S:2}
The patient cohort utilized in this challenge was on average 10.42 years old, 70.3\% male, with an average FSIQ of 111.3, and a handedness score of 0.7. Neurotypical controls in training, validation, and test data were not significantly different from their corresponding patient population in terms of sex, age, FSIQ, and handedness (all $p>0.05$).

The evaluation of the performance for all participants and datasets is summarized in Table~\ref{tab:performance}. Using the validation set performance as a baseline for the Primary Classification Task~I (ADHD), we observed a large drop in performance for participants S1, S3, and S4, across all metrics against the test set. The performance of the methods submitted by participants S2 and S5, however, remained relatively stable. For the Transference Classification Task~II (ASD with ADHD comorbidity), we see a further decrease in performance compared with Task~I, across all metrics and for all participants. 

\begin{sidewaystable}[htbp]
    \centering
    \caption{Summary of performance for each participant (S1 to S5) given as the median of each evaluation metric for validation, Task~I, and Task~II test datasets.}
    \label{tab:performance}
    \footnotesize
\begin{tabular}{|c|c|c|c|c|c||c|c|c|c|c||c|c|c|c|c|}
\hline
& \multicolumn{10}{c||}{\textbf{Task I - ADHD}} & \multicolumn{5}{c|}{\textbf{Task II - ASD}}\\
\hline
& \multicolumn{5}{c||}{\textbf{Validation}} & \multicolumn{5}{c|}{\textbf{Test}} & \multicolumn{5}{c|}{\textbf{Test}}\\
\hline
  & \textbf{S1} & \textbf{S2} & \textbf{S3} & \textbf{S4} & \textbf{S5} & \textbf{S1} & \textbf{S2} & \textbf{S3} & \textbf{S4} & \textbf{S5} & \textbf{S1} & \textbf{S2} & \textbf{S3} & \textbf{S4} & \textbf{S5} \\
  \hline
\textbf{Acc} & 0.75 & 0.53 & 0.73 & 0.83 & 0.68 & 0.5 & 0.55 & 0.53 & 0.55 & 0.68 & 0.45 & 0.53 & 0.45 & 0.43 & 0.53 \\
\hline
\textbf{AUC} & 0.73 & 0.62 & 0.85 & 0.89 & 0.72 & 0.48 & 0.63 & 0.54 & 0.47 & 0.66 & 0.45 & 0.48 & 0.43 & 0.41 & 0.56 \\
\hline
\textbf{F1} & 0.79 & 0.56 & 0.73 & 0.82 & 0.68 & 0.6 & 0.57 & 0.51 & 0.54 & 0.7 & 0.55 & 0.54 & 0.43 & 0.41 & 0.5 \\
\hline
\textbf{FDR} & 0.32 & 0.48 & 0.29 & 0.16 & 0.33 & 0.5 & 0.45 & 0.48 & 0.44 & 0.33 & 0.54 & 0.48 & 0.55 & 0.58 & 0.47 \\
\hline
\textbf{FNR} & 0.05 & 0.4 & 0.25 & 0.2 & 0.3 & 0.25 & 0.4 & 0.5 & 0.5 & 0.3 & 0.3 & 0.45 & 0.6 & 0.6 & 0.5 \\
\hline
\textbf{FOR} & 0.08 & 0.47 & 0.26 & 0.19 & 0.32 & 0.5 & 0.44 & 0.48 & 0.45 & 0.3 & 0.62 & 0.47 & 0.54 & 0.58 & 0.48 \\
\hline
\textbf{FPR} & 0.45 & 0.55 & 0.3 & 0.15 & 0.35 & 0.75 & 0.5 & 0.5 & 0.4 & 0.35 & 0.8 & 0.5 & 0.5 & 0.55 & 0.45 \\
\hline
\textbf{GM} & 0.8 & 0.56 & 0.73 & 0.82 & 0.68 & 0.61 & 0.57 & 0.51 & 0.54 & 0.7 & 0.56 & 0.54 & 0.43 & 0.41 & 0.5 \\
\hline
\textbf{Inf} & 0.5 & 0.05 & 0.45 & 0.65 & 0.35 & 0.00 & 0.1 & 0.05 & 0.1 & 0.35 & -0.1 & 0.05 & -0.1 & -0.15 & 0.05 \\
\hline
\textbf{Mark} & 0.6 & 0.05 & 0.45 & 0.65 & 0.35 & 0.00 & 0.1 & 0.05 & 0.1 & 0.36 & -0.16 & 0.05 & -0.1 & -0.15 & 0.05 \\
\hline
\textbf{MCC} & 0.55 & 0.05 & 0.45 & 0.65 & 0.35 & 0.00 & 0.1 & 0.05 & 0.1 & 0.35 & -0.13 & 0.05 & -0.1 & -0.15 & 0.05 \\
\hline
\textbf{NPV} & 0.92 & 0.53 & 0.74 & 0.81 & 0.68 & 0.5 & 0.56 & 0.52 & 0.55 & 0.7 & 0.39 & 0.53 & 0.46 & 0.42 & 0.52 \\
\hline
\textbf{OP} & 29.73 & 20.86 & 28.97 & 32.97 & 26.96 & 19.5 & 22 & 20.95 & 22 & 26.96 & 17.33 & 20.86 & 18 & 16.94 & 20.86 \\
\hline
\textbf{Pre} & 0.68 & 0.52 & 0.71 & 0.84 & 0.67 & 0.5 & 0.55 & 0.52 & 0.56 & 0.67 & 0.46 & 0.52 & 0.45 & 0.42 & 0.53 \\
\hline
\textbf{Sen} & 0.95 & 0.6 & 0.75 & 0.8 & 0.7 & 0.75 & 0.6 & 0.5 & 0.5 & 0.7 & 0.7 & 0.55 & 0.4 & 0.4 & 0.5 \\
\hline
\textbf{Spec} & 0.55 & 0.45 & 0.7 & 0.85 & 0.65 & 0.25 & 0.5 & 0.5 & 0.6 & 0.65 & 0.2 & 0.5 & 0.5 & 0.45 & 0.55 \\
\hline
\end{tabular}%
\end{sidewaystable}
\clearpage

\begin{figure}[ht!]
    \centering
    \includegraphics[width=.9\linewidth]{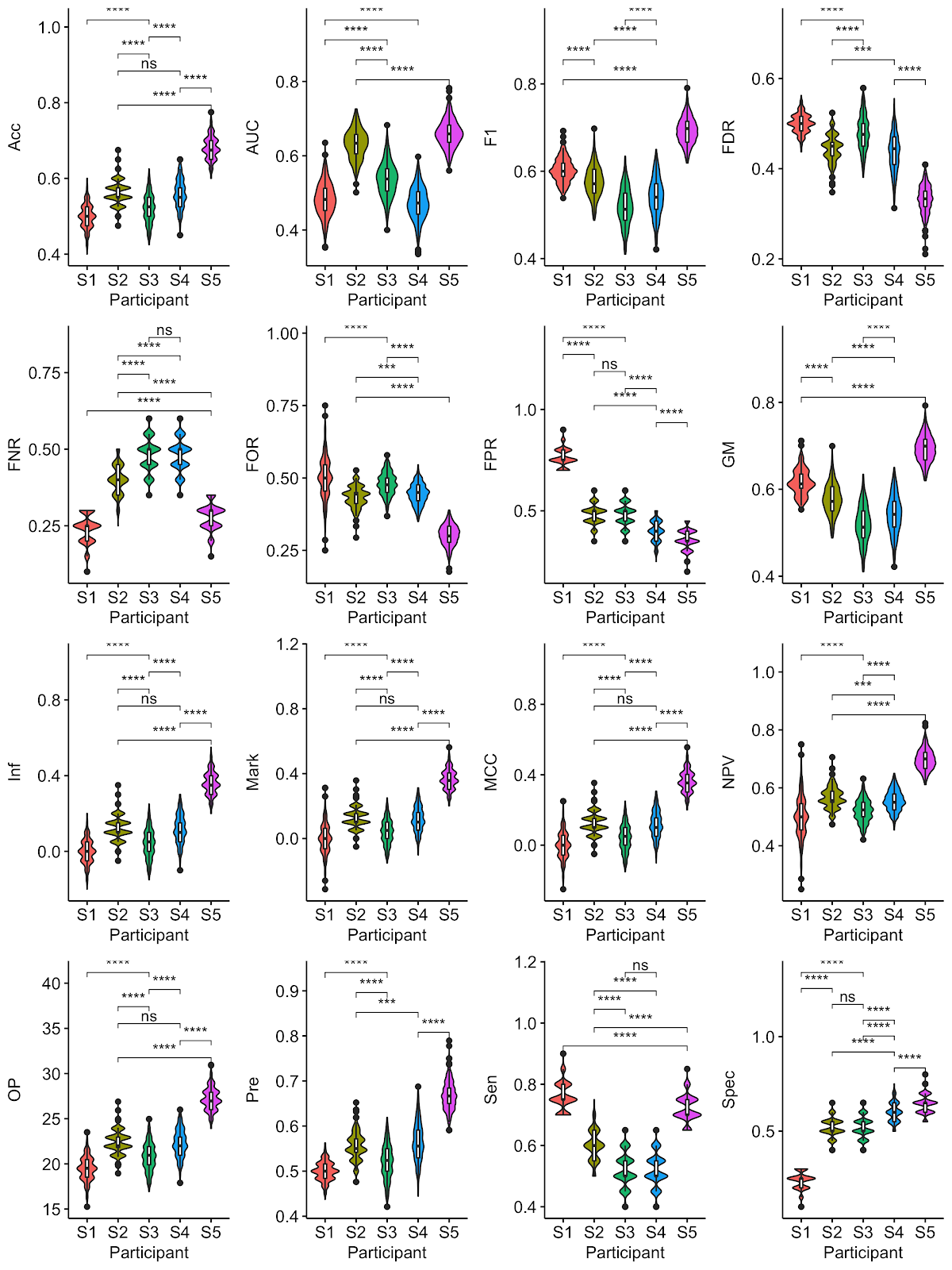}
    \caption{Primary Classification Task I (ADHD versus NC) evaluation measure distributions for each participant. Statistical significance was determined based on pair Wilcoxon test (ns: $p>0.05$; *: $p<0.05$; **: $p<0.01$; ***: $p<0.001$; ****: $p<0.0001$), with a Bonferroni-corrected level of significance at $p \leq 0.0001$.}
    \label{fig:adhd_results}
\end{figure}

\begin{figure}[ht!]
    \centering
    \includegraphics[width=.8\linewidth]{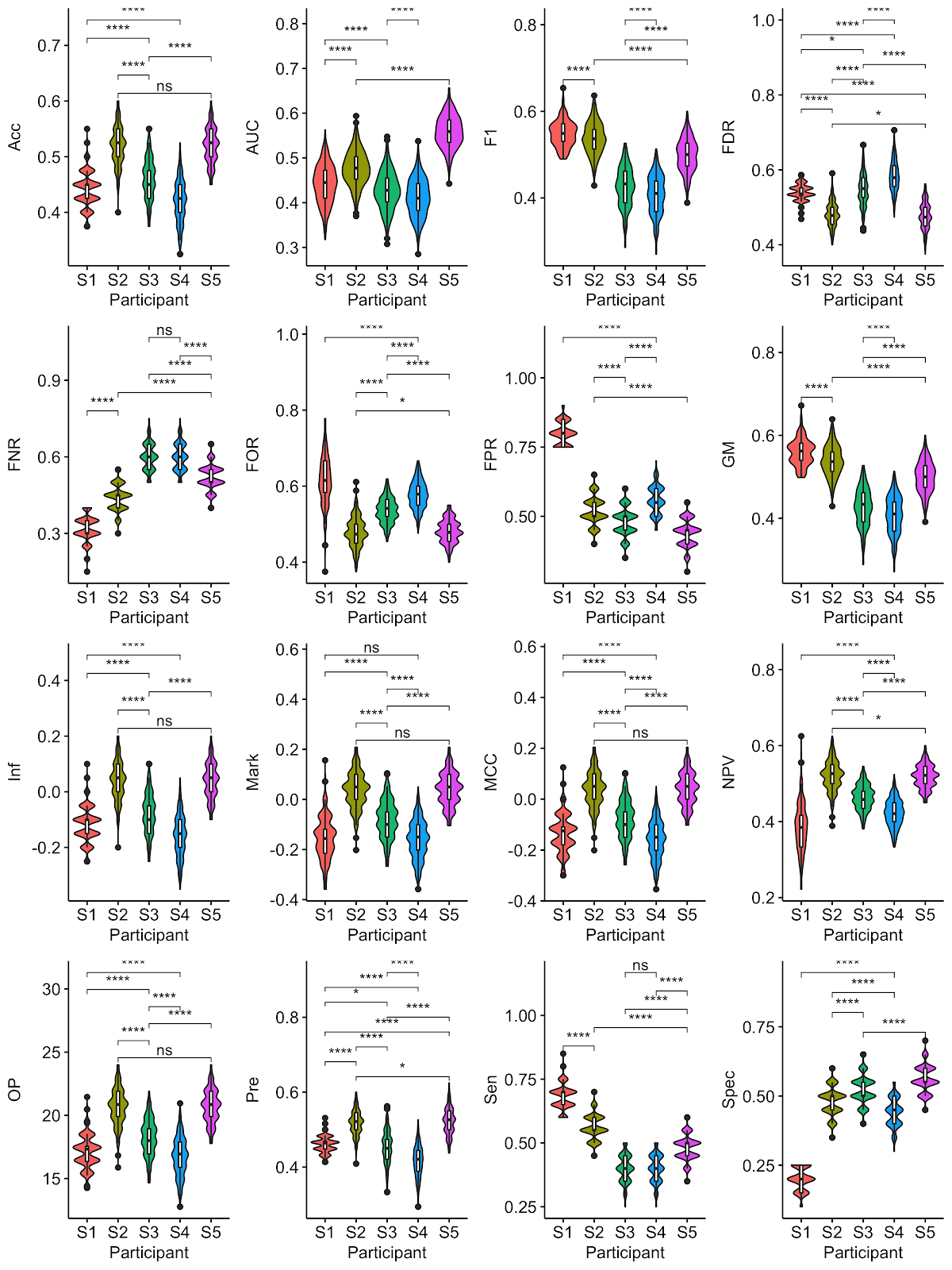}
        \caption{Transference Classification Task II (ASD versus NC) evaluation measure distributions for each participant. Statistical significance was determined based on pair Wilcoxon test (ns: $p>0.05$; *: $p<0.05$; **: $p<0.01$; ***: $p<0.001$; ****: $p<0.0001$), with a Bonferroni-corrected level of significance at $p \leq 0.0001$.}
    \label{fig:hfa_results}
\end{figure}

Figures~\ref{fig:adhd_results} and~\ref{fig:hfa_results} illustrate the distribution of each evaluation metric for each participant's model on Task~I and Task~II, respectively. Each subplot indicates which of the submissions performs best with respect to a specific metric, as described in Table~\ref{tab:measures}. The performance evaluation was based on the cross-validation setup and significance between ranks are delineated.

Table~\ref{tab:ranking} summarizes the participant rankings based on the results of the metrics presented in Figure~\ref{fig:adhd_results} and~\ref{fig:hfa_results}, including the best median for each metric (minimum or maximum) across participants. The overall ranking for Task I between ADHD and NC is as follows (participant (rank product)): S5 (1.1), S2 (2.4), S4 (2.7), S1 (3.1), and S3 (3.6). The overall ranking for the Transference Classification Task II between ASD and NC is as follows: S5 (1.3), S2 (1.4), S1 (2.3), S3 (2.6), and S4 (3.6). In both Tasks, S5 and S2 consistently ranked first and second, respectively, in our evaluation. 

\bgroup
\linespread{1.2}
\begin{table}[ht!]
    \centering
    \caption{Rankings of each participant for each measure and both classification tasks, including the best median for each metric across classifiers for ADHD and ASD test cohorts.}
    \label{tab:ranking}
    \tiny
    \begin{tabular}{|c|c|c|c|c|c||c|c|c|c|c||c|c|}
    \hline
        & \multicolumn{10}{c||}{\textbf{Rankings}} & \multicolumn{2}{c|}{\textbf{Best median metric}} \\ \hline 
        & \multicolumn{5}{c||}{\textbf{Task I - ADHD}} & \multicolumn{5}{c||}{\textbf{Task II - ASD}} & \multirow{2}{*}{\textbf{ADHD}} & \multirow{2}{*}{\textbf{ASD}} \\ \cline{1-11}
         & \textbf{\#1} & \textbf{\#2} & \textbf{\#3} & \textbf{\#4} & \textbf{\#5} & \textbf{\#1} & \textbf{\#2} & \textbf{\#3} & \textbf{\#4} & \textbf{\#5} &&\\ \hline 
        \textbf{Acc} & S5 & \multicolumn{2}{c|}{S2/S4} & S3 & S1 & \multicolumn{2}{c|}{S2/S5} & S3 & S1 & S4 & 0.68 & 0.53 \\ \hline 
        \textbf{AUC} & S5 & S2 & S3 & S1 & S4 & S5 & S2 & S1 & S3 & S4 & 0.66 & 0.56 \\ \hline 
        \textbf{F1} & S5 & S1 & S2 & S4 & S3 & S1 & S2 & S5 & S3 & S4 & 0.70 & 0.55 \\ \hline 
        \textbf{FDR} & S5 & S4 & S2 & S3 & S1 & \multicolumn{2}{c|}{S2/S5} & \multicolumn{2}{c|}{S1/S3} & S4 & 0.33 & 0.47 \\ \hline 
        \textbf{FNR} & S1 & S5 & S2 & \multicolumn{2}{c||}{S3/S4} & S1 & S2 & S5 & \multicolumn{2}{c||}{S3/S4} & 0.25 & 0.30 \\ \hline 
        \textbf{FOR} & S5 & S2 & S4 & S3 & S1 & \multicolumn{2}{c|}{S2/S5} & S3 & S4 & S1 & 0.30 & 0.47 \\ \hline 
        \textbf{FPR} & S5 & S4 & \multicolumn{2}{c|}{S2/S3} & S1 & S5 & S2 & S3 & S4 & S1 & 0.35 & 0.45 \\ \hline 
        \textbf{GM} & S5 & S1 & S2 & S4 & S3 & S1 & S2 & S5 & S3 & S4 & 0.70 & 0.56 \\ \hline 
        \textbf{Inf} & S5 & \multicolumn{2}{c|}{S2/S4} & S3 & S1 & \multicolumn{2}{c|}{S2/S5} & S3 & S1 & S4 & 0.35 & 0.05 \\ \hline 
        \textbf{Mark} & S5 & \multicolumn{2}{c|}{S2/S4} & S3 & S1 & \multicolumn{2}{c|}{S2/S5} & S3 & \multicolumn{2}{c||}{S1/S4} & 0.36 & 0.05 \\ \hline 
        \textbf{MCC} & S5 & \multicolumn{2}{c|}{S2/S4} & S3 & S1 & \multicolumn{2}{c|}{S2/S5} & S3 & \multicolumn{2}{c||}{S1/S4} & 0.35 & 0.05 \\ \hline 
        \textbf{NPV} & S5 & S2 & S4 & S3 & S1 & \multicolumn{2}{c|}{S2/S5} & S3 & S4 & S1 & 0.70 & 0.53 \\ \hline 
        \textbf{OP} & S5 & \multicolumn{2}{c|}{S2/S4} & S3 & S1 & \multicolumn{2}{c|}{S2/S5} & S3 & \multicolumn{2}{c||}{S1/S4} & 26.96 & 20.86 \\ \hline 
        \textbf{Pre} & S5 & S4 & S2 & S3 & S1 & \multicolumn{2}{c||}{S2/S5} & \multicolumn{2}{c|}{S1/S3} & S4 & 0.67 & 0.53 \\ \hline 
        \textbf{Sen} & S1 & S5 & S2 & \multicolumn{2}{c||}{S3/S4} & S1 & S2 & S5 & \multicolumn{2}{c||}{S3/S4} & 0.75 & 0.70 \\ \hline 
        \textbf{Spec} & S5 & S4 & \multicolumn{2}{c|}{S2/S3} & S1 & S5 & S3 & S2 & S4 & S1 & 0.65 & 0.55 \\ \hline 
    \end{tabular}
\end{table}
\egroup

\begin{figure}[ht!]
    \centering
    \includegraphics[width=.9\linewidth]{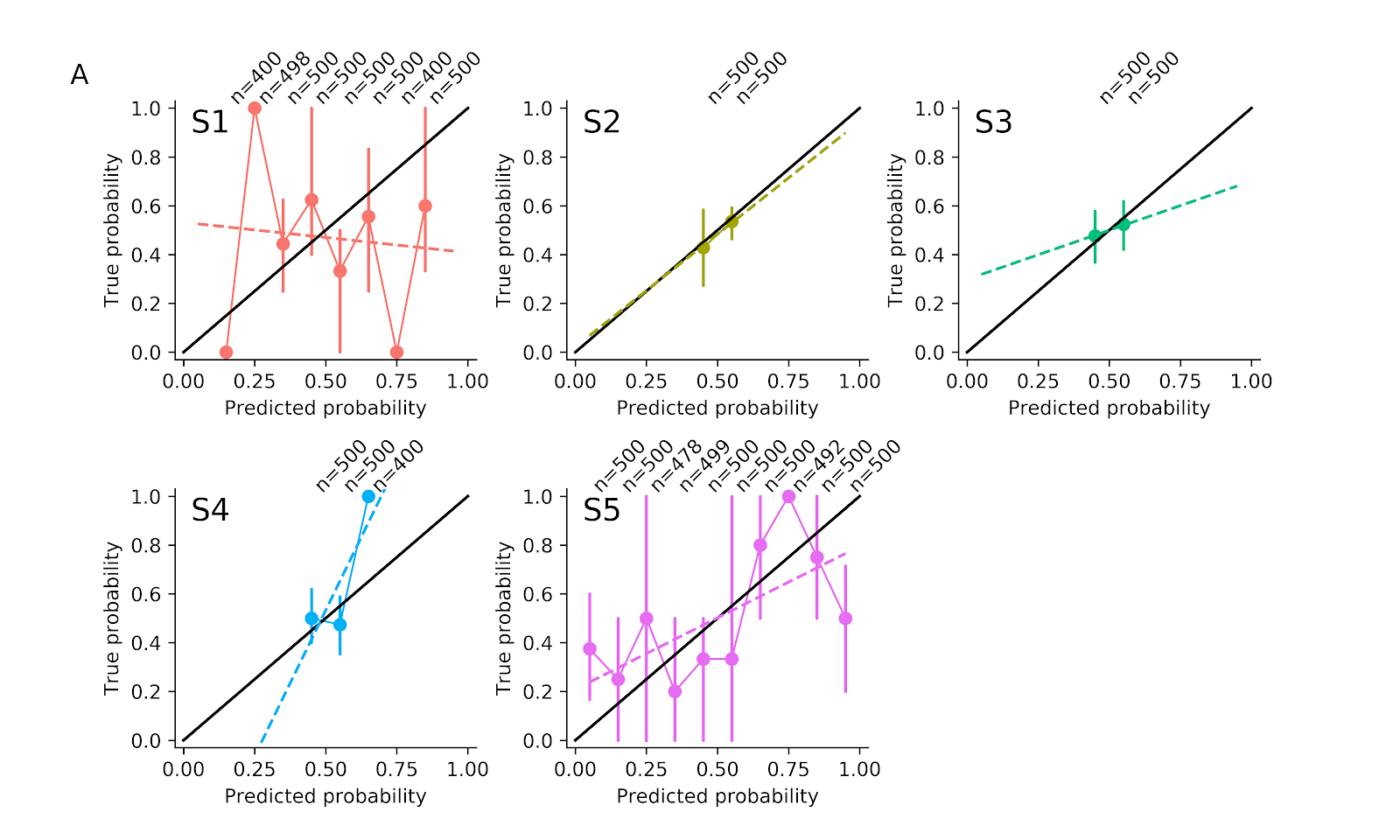}\\
    \includegraphics[width=.9\linewidth]{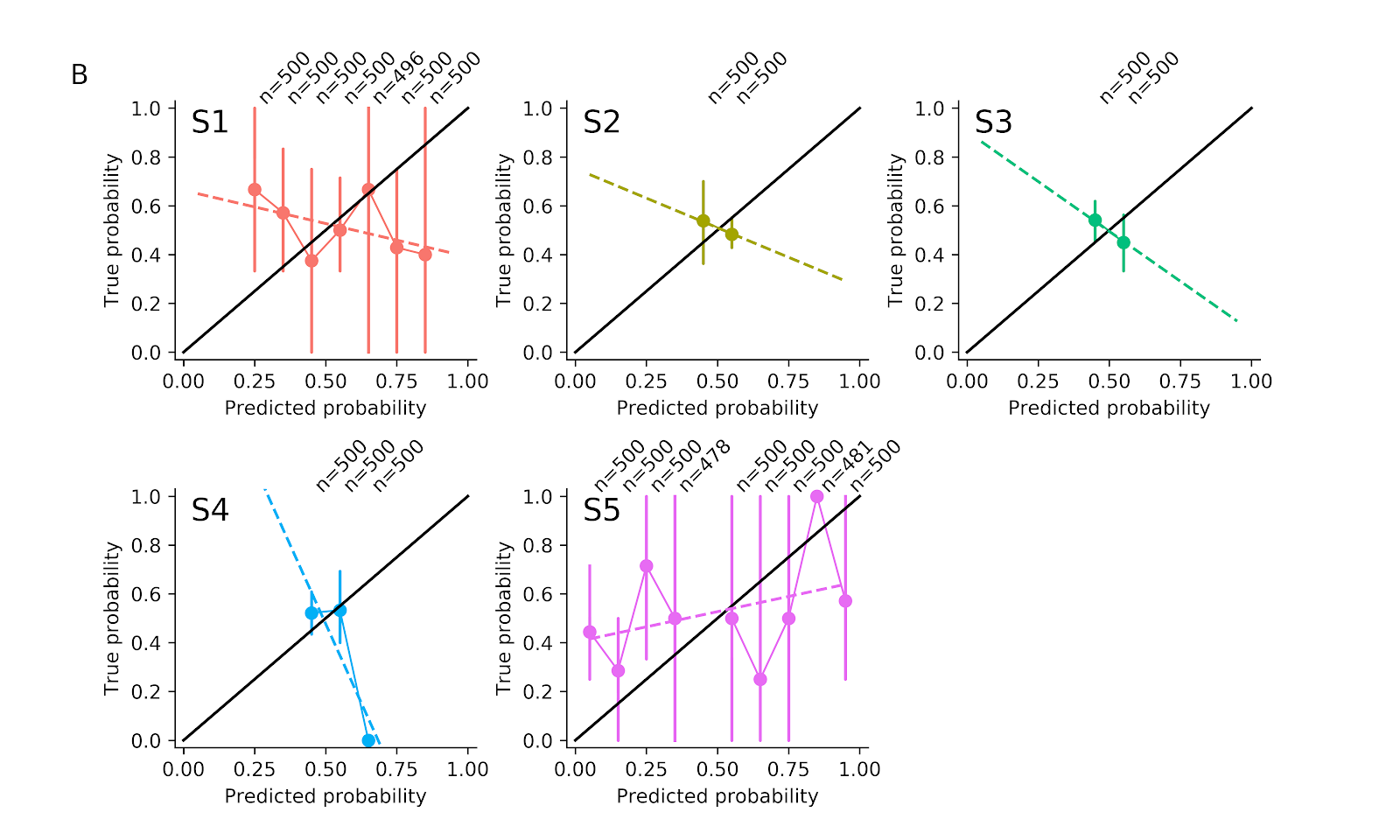}    
    \caption{Calibration curves for classification A) Task I (ADHD) and B) Task II (ASD). The {\textit{n}} indicates the number of subsets that contributed to the distribution within each bin. Dashed lines represent the resulting linear regression from the data. }
    \label{fig:calibration}
\end{figure}

Figure~\ref{fig:calibration} shows the calibration curve for each submission and classification task. A good classification model is represented by a sigmoid or step function. Generally, we observe an expected positive trend, where subjects with higher probability scores are more likely to be ADHD patients in Task~I (Figure~\ref{fig:calibration}A). For classifying ASD patients with ADHD comorbidity, most methodologies predominantly assigned higher probabilities to controls (Figure~\ref{fig:calibration}B). Additionally, out of all submissions, three did not use the entire predicted probability spectrum (S2, S3, S4), with a reversal in their linear fit between Task~I and Task~II. 

\section{Discussion}
In this paper, we have described the setup, standardized assessment, and results of the first Connectomics in Neuroimaging Transfer Learning Challenge, hosted at the 22nd International Conference on Medical Image Computation and Computer Assisted Intervention (MICCAI) 2019 in Shenzhen, China. The CNI-TL Challenge was designed to probe the generalizability and clinical relevance of classification methodologies, which are now growing in popularity for functional connectivity data.

CNI-TLC combines two developmental disorders, ADHD and ASD, which individually have complex yet distinct behavioral phenotypes and diagnostic assessments. At the same time, the co-occurrence of these disorders is high with many ASD children also exhibiting the stereotypical attention problems and impulsivity of ADHD. In fact, connectomics studies have combined ASD and ADHD populations to disentangle their joint clinical presentations, with findings of both shared and distinct functional network features between ADHD and ASD (\cite{Ray2014,diMartino2013,lake2019}). Our unique setup extends the conventional notion of ``transfer learning'' to further probe this phenomenon. Rather than transferring just the model architecture optimized for one cohort and re-training it on a second cohort, we transfer the learned representations themselves. In this manner, we test whether a robust symptomatology can be learned for ADHD, and if so, whether it can also be extracted in a co-morbid population. Furthermore, our Challenge setup is in line with recent trends in computational neuroscience to develop methods that learn from multiple populations, so as to increase their clinical relevance (\cite{Douw2019}).

\subsection{Primary Classification Task I - ADHD versus NC}
No method performed best nor worst on all metrics as shown in Table~\ref{tab:ranking}. However, in three out of five submissions, we observed a significant drop in performance between validation and test data, whereas methods S2 and S5 were able to retain their performance. A drop when moving from seen to unseen data in the evaluated metrics is to be expected, however, considering the even split between patients and controls, many of the methods generalized to near-chance classification accuracy. Furthermore, the validation and test datasets were matched to the training data on all phenotypic variables. Hence, Table~\ref{tab:ranking} suggests an overfitting to the released data.

Erring on the side of caution, frameworks intended for use in the clinic will often prefer to optimize on FDR (type I error) and FNR (type II error), instead of aiming to perform unilaterily well in all the presented metrics. If we were to assess performance by this criteria, Table~\ref{tab:performance} shows S5 outperforming other methods on the unseen ADHD test data (average of FDR and FNR: S1: 0.38; S2: 0.43; S3: 0.49; S4: 0.47; S5: 0.32). While S5 demonstrates the best average value, a FDR and FNR of 0.3 can still be seen as too high, where 30\% of the patients and controls are misclassified.

We can further differentiate the methodologies based on the results of the calibration curves. Three out of five classifiers (S2, S3 and S4) produced a very narrow range of probabilities for classification (around $0.5 \pm 0.1$), whereas the two remaining classifiers utilized (almost) the full probability spectrum. Not utilizing the full predicted probability spectrum may reflect the models' uncertainty in classification and could potentially lead to misclassification in case of ``noise'' in the data. Thus, S5 demonstrated the overall best performance on unseen data. Furthermore, the calibration curve for S5 vaguely resembles a sigmoidal shape with an inflection point around a predicted probability of 0.6.

\subsection{Transference Classification Task II - ASD with ADHD Comorbidity versus NC}
Only one method (S5) exhibited the correct trend of predicted probabilities in classifying ASD patients with ADHD comorbidity (Figure~\ref{fig:calibration}B). All other methodologies predominantly assigned a higher probability of disease for NC, than for ASD patients. This may highlight the difficulties of achieving generalizability in these models if other comorbidities are present, and further agrees with our quantitative ranking of methodologies.

Overall, the large decrease in performances between Task~I and Task~II (Table~\ref{tab:performance}), along with differences in the calibration curves (Figure~\ref{fig:calibration}) have two likely explanations. First, the models are overfitting to the released datasets. Second, the predictive rsfMRI features learned for ADHD do not transfer onto the ASD cohort with an ADHD co-morbidity, either because the latter has a unique neural phenotype, or because the commonalities between the disorders are overwhelmed by other signatures in the rsfMRI data, which did not allow the extraction of biologically meaningful information that could be specific to ADHD. 

\subsection{Approaches and Considerations for Future Challengers}
While we cannot state unequivocally one submission that performed the best across all evaluation metrics in both Tasks (see Table~\ref{tab:performance}), we can draw inferences from our submissions that may help to improve future developments of connectomic classification methodologies.

Connectivity matrix estimation varies across rsfMRI connectome studies and has been shown to affect classification performance (\cite{abraham2017deriving}). This lack of consensus in the field is reflected in our submissions - with each method using a different type of connectome as input. The same observation can be made for choice of atlas, which can also influence performance (\cite{abraham2017deriving}). Interestingly, none of the CNI-TLC submissions used traditional graph metrics (\cite{Rubinov}), such as node degree, centrality, and small-worldness, which are ubiquitous in functional connectivity literature. Given that machine-/deep-learning methods can utilize big data during training, utilizing connectome data `as is' seems intuitive, and it may be that traditional graph metrics are not favored for collapsing the number of data points in a time series. Even so, there may be advantages to including metrics that characterize complex global and topological properties unique to brain networks which may not be directly interpreted from raw connectivity matrices or timeseries data.

Despite the growing popularity of deep-learning methods, only one submission used an end-to-end deep-learning framework. Instead, the most common classifier among Challenge submissions was the support vector machine (\cite{Cortes}). Ultimately, the deep-learning approach (S5) had the highest overall rank. Part of the reason may be attributed to an initial feature selection being performed and using only the most discriminative connections as input to the classifier. In addition, they employed a dynamic model that can capture key temporal information in the rsfMRI signal. The second place submission on classification Task I, in contrast to others, relied on the normalized graph Laplacian (S2). The graph Laplacian and its eigenspectrum enable a mapping of discrete data (a network) into vector spaces and manifolds, and their advantages have been greatly investigated, e.g., in fields such as clustering (\cite{chung_spectral_1997}). This form of network representation has also found its way into connectomes and may have potential to capture further discriminative information (\cite{abdelnour_network_2014,CHUNGNeuroImg2016, ChungPRNI2016, SchirmerChungCNI19}). These data filtering or manipulation techniques are effective for their robustness to overfitting (\cite{Du2018}) and may be a reason for S5 and S2 achieving the best generalization performance from Task~I to Task~II. 

We also observed that not all available information was utilized by participants. While the winning algorithm pre-selected the most discriminating features, this approach could be extended to feature selection across multiple atlases. This observation highlights the importance of including prior knowledge in developing classifiers for clinical tasks. Generally, we observed that most of the classifiers did not cover the full predicted probability spectrum. While this calibration may not be directly related to a classifier's performance on the actual task, it can serve as a proxy for assessing whether an algorithm is extracting meaningful information from the data and can reveal issues in the classification process. 

Another option for future challenges is the pooling of all classification results to asses their performance as an aggregate entity. As participants were asked to output classification scores, we combined all methodologies \textit{post hoc} by obtaining their collective average, median, and maximum confidence (from the predicted value that is furthest away from 0.5), see~\ref{app:C}, Table~\ref{tab:appendix1}). Subsequently, if a consensus vote outperforms the individual methodologies, secondary analyses are required to fully understand the implications.

\subsection{Challenge Contributions}
The data used for this Challenge represents one of the largest collections of rsfMRI data acquired at a single site for three different cohorts (ADHD, ASD, and NC). The data was carefully and consistently collected on a single scanner model (Philips Achieva) by researchers at KKI. The in-house preprocessing pipeline is also standard in the field and included intermediate manual checks for data quality. 

As part of our aim was to investigate the classification performance of several methodologies, we chose to release data that was uniformly processed by a single pipeline that has been previously published (see ~\cite{MUSCHELLI201422,NEBEL2016633,stoodley2017}).
The effect of rsfMRI processing decisions on analysis is a large and important issue to address that is beyond the scope of our Challenge, but has been investigated by others (see e.g. ~\cite{Bowring2019, Vytvarova2017,Vergara2017}). We note that a subset of the rsfMRI data has been released through ABIDE (\cite{ABIDE}) and ADHD-200\footnote{http://fcon\_1000.projects.nitrc.org/indi/adhd200/results.html}. However, we were careful to ensure that the testing set contained only private data that had never been released and, as only pre-trained models were accepted, none of the algorithms were retrained with access to the test data. Each of our training, validation, and testing datasets were matched on demographic variables, in order to remove extraneous confounds. Finally, we opted to release the average time series from the rsfMRI data, rather than static or dynamic connectivity matrices, to provide more flexibility for participants. Likewise, we used three popular atlases to provide a range of spatial resolutions. 

In order to standardize our evaluation, we implemented a novel framework for participants to submit pre-trained models as Docker images. The resulting Docker image from our framework also doubles as a plug-in compatible with the ChRIS platform\footnote{\url{https://chrisproject.org}}, a pervasively open source framework that utilizes cloud technologies to democratize medical analytics application development. ChRIS allows researchers the ability to simply deploy the same application they have already developed in a cloud infrastructure with access to more data, more computational resources, and more collaboration to drive medical innovation, while standardizing healthcare application development. Subsequently, our approach gives participants the option to share their pre-trained model with the wider medical imaging research community. 

One novel aspect of our Challenge, in comparison to current practices, is the large array of evaluation metrics used to assess the performance of each algorithm. This decision was motivated by a lack of consensus in the field about which metric is the ``best'' for evaluation, as it can be seen through the plethora of assessment metrics (\cite{hossin2015review}). Importantly, each metric assesses different aspects of an algorithm's performance. Another key strength of our evaluation procedure is that the test data is kept private. The unseen test data allows us to probe model overfitting and provides an objective comparison between algorithms. 

With the above contributions, CNI-TLC has maintained and followed guidelines specifically outlined for challenges advocated by MICCAI with the intention of upholding transparency and fairness (\cite{maier-hein2018rankcomp}).

\subsection{Future Work}

One of the main takeaways of our CNI-TL Challenge is that significant improvements are necessary in order to translate functional connectivity into clinical practice. Considering the performance across all participants summarized in Table~\ref{tab:ranking}, the overall scores are relatively low, with even the best results on the ASD cohort closely resembling chance in most metrics. 

The CNI-TL Challenge highlights some important recommendations for future work on classification using functional connectivity. For example, our Challenge demonstrates the well-known issue of model generalizability to unseen data. This result supports the general recommendation that the data should be split into training, validation, and testing sets, with the testing set evaluated only once at the end of the study. Accordingly, we withheld the test set from CNI-TLC participants and continue to do so from the public, allowing the continuation of the Challenge, while methodologies can be further evaluated through the ChRIS platform.

By pooling submissions to solve the same problem, our Challenge has demonstrated the diversity and creativity of methods developed to use functional connectomes for classification. With specifics to model- and neural network-based methods, CNI-TLC further highlighted the need for a consensus on a series of experimental design tests to understand the influence of data representation (i.e. matrix estimation type, atlas, thresholding) in relation to an approach, disease, or dataset. Given the huge effort in recent years to amass and coalesce data from multiple sites, perhaps a similar ethos of combining and sharing efforts and approaches is equally necessary to tackle the neuroimaging challenges of today.

In recognition of the above points and of the role that challenges play in the field, we will keep the CNI-TL Challenge available and online for scientists to continually test their approaches. Submission is open to everyone, whether or not they have participated in the Challenge. Specifically, we provide a website\footnote{\url{https://fnndsc.childrens.harvard.edu/cnichallenge}} for developers of new methods to upload new solutions as ChRIS plug-ins, with an automated evaluation infrastructure ``behind the scenes'' offering dynamic feedback on the performance of their model on the test set. Importantly, the test data used is not exposed, thus providing a fair baseline from which to measure comparative performance.

\section{Conclusion}

The CNI-TL Challenge is an important step in the field of connectomics. First, it demonstrates the necessity of objective evaluation to assess generalizability to unseen data. Second, it goes beyond a single task-optimization setting to address the key question of whether we are capturing biologically meaningful phenomena. Here, we showed that the classification performance of all methods dropped nearly to chance on a patient population with the target disease as a comorbidity. This result underscores the need for further work to reach clinical translation of functional connectomics for disease identification. With the training and validation data remaining publicly available, and the test data accessible through an online evaluation platform, CNI-TLC facilitates continual development of new classification methods for connectomics.

\section*{Acknowledgements and Disclaimers}

\underline{Organizers:}

M. D. Schirmer was supported by the European Union’s Horizon 2020 research and innovation programme under the Marie Sklodowska-Curie grant agreement No 753896.

A. Venkataraman was supported by the National Science Foundation Collaborative Research in Computational Neuroscience (CRCNS) award 1822575 and the National Science Foundation CAREER award 1845430.

I. Rekik is supported by the 2232 International Fellowship for Outstanding Researchers Program of TUBITAK (Project No:118C288) and the the European Union's Horizon 2020 research and innovation programme under the Marie Sklodowska-Curie grant agreement No 101003403.

M. Kim was supported by UNC Greensboro New Faculty Award.

A. W. Chung was supported by the American Heart Association and Children’s Heart Foundation Congenital Heart Defect Research Award, \\*19POST34380005.

\underline{Data team:} 

The patient recruitment, data acquisition, and preprocessing was supported by the Autism Speaks Foundation (awards 1739 and 2384) and by the National Institutes of Health under the following awards: K02 NS44850 (PI Mostofsky), R01 MH078160-10 (PI Mostofsky), R01 MH085328-09 (PI Mostofsky), R01 MH106564-03 (PI Edden), K23 MH101322-05 (PI Rosch), K23 MH107734-05 (PI Seymour), R01 NS048527-08 (PI Mostofsky), R01 NS096207-05 (PI Mostofsky), R01MH085328-14 (PI Mostofsky), U54HD079123, UL RR025005, and P54 EB15909.

This work was prepared while Karen Seymour was employed at Johns Hopkins University and Kennedy Krieger Institute. The opinions expressed in this article are the author's own and do not reflect the view of the National Institutes of Health, the Department of Health and Human Services, or the United States government.   

\underline{Participants:}

MeInternational: This work was supported by the EPSRC-funded UCL Centre for Doctoral Training in Medical Imaging (EP/L016478/1); the National Institute for Health Research (NIHR); the Wellcome Trust (210182/Z/18/Z, 101957/Z/13/Z) and the Medical Research Council UK (Ref MR/J01107X/1). 

HSE: This work was supported by the Russian Academic Excellence Project `5-100'.

ShefML: This work was supported by grants from the UK Engineering and Physical Sciences Research Council (EP/R014507/1).

YaleIPAG: N. C. Dvornek was supported by the National Institute of Health (NIH grants R01MH100028 and R01NS03519). J. Zhuang was supported by the National Institute of Health (NIH grant R01NS03519).

\appendix
\section{}
\label{app:A}
Data used in this challenge were drawn from retrospective data acquired at the Kennedy Krieger Institute. The associated study names and IRB approval numbers are as follows:

\begin{enumerate}
    \item Neurologic Basis of Inhibitory Deficits in ADHD (IRB 02-11-25-01)
    \item Anomalous Motor Physiology in ADHD (IRB NA\_00000292)
    \item Deficient Response Control in ADHD (IRB NA\_00027428)
    \item Edden MRS (IRB NA\_00088856)
    \item Rosch Delay Discounting in ADHD (IRB 00032351)
    \item Seymour Neurobehavioral Correlates of Frustration in ADHD (IRB 00063119)
    \item Motor Skill Learning in Autism (IRB 03-05-27-10)
    \item Motor Skill Learning in Autism: Assessment and Treatment of Altered Patterns of Learning (IRB NA\_00027073)
    \item Tactile Adaptation in Tourette's Syndrome (IRB NA\_00090977)
\end{enumerate}

\section{}
\label{app:B}
Detailed description of the methodology used by each submission.

\subsection*{Submission (S1) - MeInternational - Linear Support Vector Machine Framework to Predict Abnormal Functional Connectivity}
Team: H. Irzan, M. Hütel, S. Ourselin, N. Marlow, A. Melbourne

\begin{figure}[ht]
    \centering
    \includegraphics[width=\linewidth]{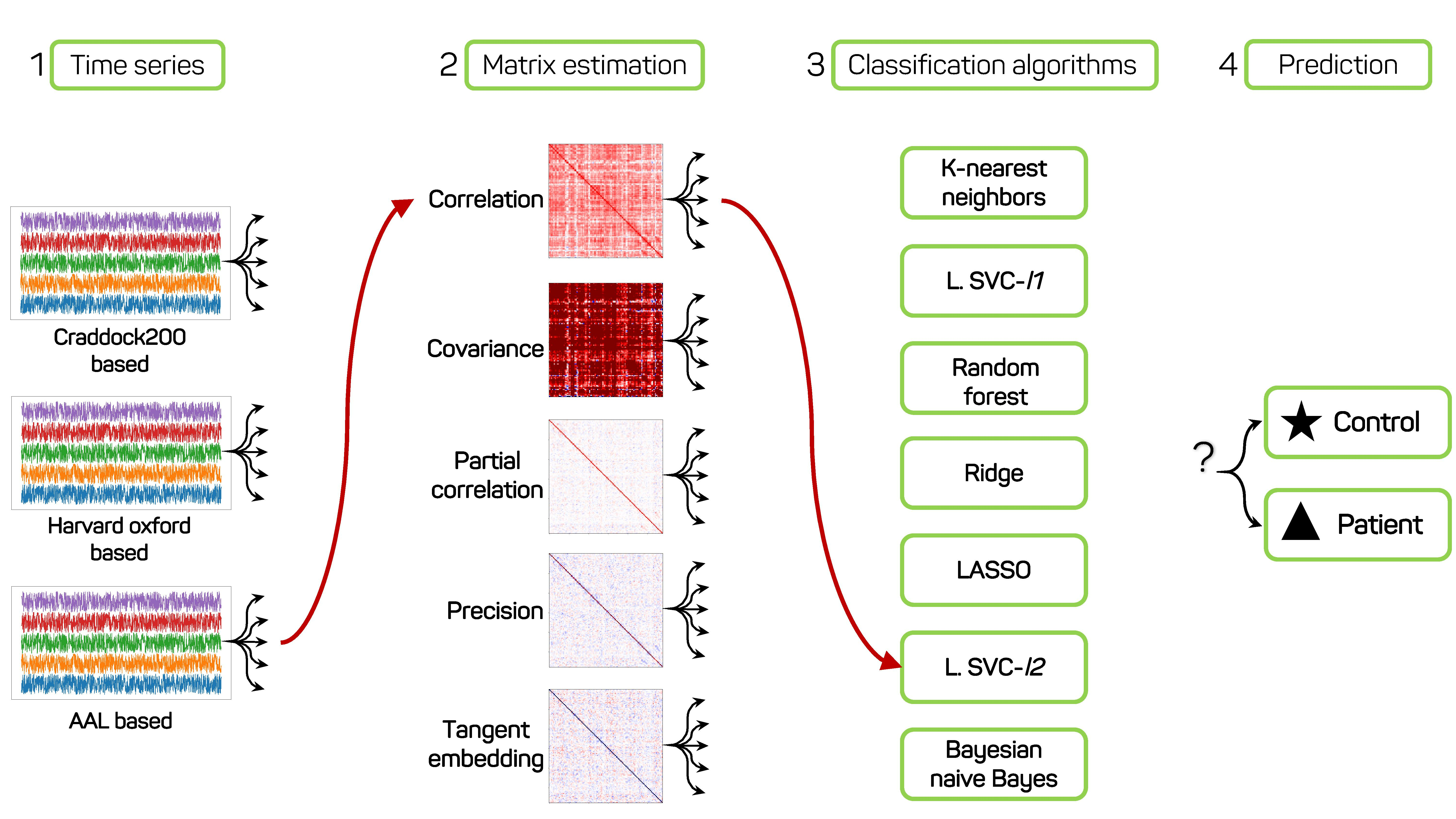}
    \caption{Classification pipeline. In step 1, time series are produced by extracting the mean time courses of the preprocessed rs-fMRI using three anatomical atlases (Craddock200, Harvard Oxford, and AAL). These were provided as part of the challenge. In step 2, the time series are transformed into functional connectomes by using five matrix estimation methods. The functional connectomes are employed to perform supervised learning task with seven classifiers in step 3 and perform class prediction in step 4. The pipeline highlighted in red demonstrated the best classification accuracy on the test set.}
    \label{fig:S1pipeline}
\end{figure}

The team evaluated different pipelines to select the best combination of functional connectivity metrics, atlases and classification algorithms based on prediction accuracy. The nodes of the functional connectome are the individual atlas defined ROIs. The authors utilized the three different anatomical brain atlases (Craddock200 with 200 ROIs, Harvard Oxford with 110 ROIs, and AAL with 116 ROIs) and examined correlation, covariance, partial correlation, precision, and tangent embedding as metrics for functional connectivity. The team investigated $l_1$ and $l_2$ norm SVMs, linear regression with either $l_1$ or $l_2$ regularization, random forest, $k$-nearest neighbor, and naive Bayes classifiers as classification algorithms. Pipelines were built by using permutations of functional connectivity matrices, anatomical atlas, and classification algorithms as in\cite{abraham2017deriving}. The performance of each pipeline was evaluated based on its prediction accuracy score on the validation set. Figure~\ref{fig:S1pipeline} shows the main steps of the methodology. The best performance was achieved using the AAL atlas, correlation metric, and SVM with $l_2$ regularization and penalty parameter of the error term $C=3^{-5}$, as highlighted in red in Figure~\ref{fig:S1pipeline}.

\subsection*{Submission 2 (S2) - HSE - Classification of ADHD disorder Against Healthy Based on the Spectra of Normalized Laplacians}
Team: E. Levchenko

The resting state time-varying signals of the rsfMRI data can be considered as a graph (connectome) (\cite{sporns2011human}). Here, for each subject, a correlation matrix $W_{ij}$ was calculated between each pair of ROIs $i$ and $j$, using the Pearson correlation coefficient. First, the main diagonal in $W$ and all negative values $W_{ij}<0$ were set to zero, i.e. the graph was reduced to its positive weight subgraph (see e.g.\cite{chung2019network}). 

Graph theoretical studies have widely utilized the normalized Laplacian to characterize networks (see e.g.\cite{chung1997spectral,CHUNGNeuroImg2016}). The normalized Laplacian matrix $L$ is defined as

\begin{equation}
L = D^{-1/2} (D - W)  D^{-1/2},
\end{equation}
where each node in diagonal matrix $D$ is given as $d_{i} = \sum_{j}A_{ij}$. The normalized Laplacian has several useful properties - one of them being that the eigenvalues are between 0 to 2 (\cite{chung1997spectral}). 

In this challenge, rsfMRI time series data were provided based on multiple atlases with different sizes for each subject. Here, the normalized Laplacian $L$ was calculated for each subject and atlas independently. The distributions of eigenvalues of matrix $L$ for each atlas were concatenated into one feature vector of length 426 for each subject. This feature vector serves as input for the classification problem, which was addressed using an SVM algorithm with a polynomial kernel. The hyperparameters were optimized on the provided training set and the classification accuracy was obtained on the validation set using 10-fold cross-validation.

\subsection*{Submission 3 (S3) - ShefML - Domain Independent SVM for CNI Challenge}
\label{sec:S3}
Team: S. Zhou, M. Kunda, H. Lu

\begin{figure}[ht]
    \centering
    \includegraphics[width=\linewidth]{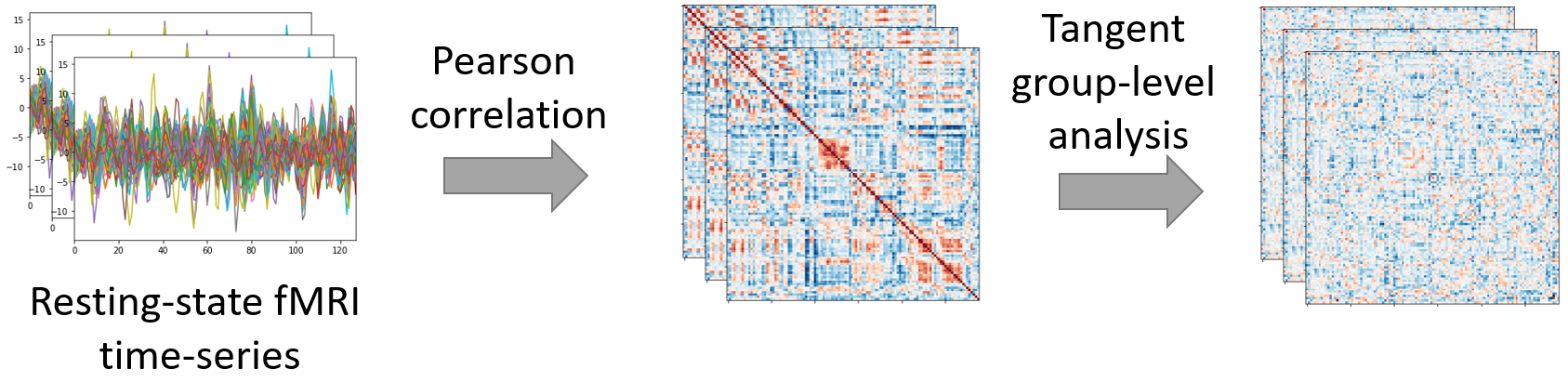}
    \caption{Extracting Tangent Pearson connectivity features from resting-state fMRI time-series.}
    \label{fig:tp}
\end{figure}

This solution learns a model via a two-stage pipeline: \textit{feature extraction} and \textit{classifier training}. For the feature extraction stage, Tangent Pearson (TP) (\cite{kunda2020improving}) is applied to extract features from resting-state time-series of the AAL atlas. Specifically, as illustrated in Fig. \ref{fig:tp}, Pearson correlation of the ROI-to-ROI relationship for each subject is computed first. Then tangent correlation analysis (\cite{varoquaux2010detection}) is performed to learn the group-level features of the brain network correlations, i.e. the connectivity of connectives. Additionally, the mean and standard deviation of each time series are concatenated to the TP features. 

For classification, it is assumed that the decisions (patient or control) made by a classifier should be independent to subjects' information, such as gender and age, and therefore Side Information Dependence Regularization (SIDeR) learning framework (\cite{zhou2020side}) is used to leverage the subjects' phenotype information for model training. The learning framework is given by

\begin{equation}
    \min_{f}  \underbrace{\mathcal{L}(f(\mathbf{X}^l), \mathbf{Y})}_{\textrm{Empirical risk}} + \underbrace{\sigma\|f\|^2_K}_{\textrm{Model complexity}} + \underbrace{\lambda \rho(f(\mathbf{X}), \mathbf{D})}_{\textrm{Side information dependence}},
    \label{eq:sider}
\end{equation}
where $\rho(\cdot, \cdot)$ denotes a statistical independence metric, $\sigma$ and $\lambda$ are hyper-parameters, $ \mathbf{X}^l $, $\mathbf{Y}$, $\mathbf{X}$, $\mathbf{D}$ denote the labelled instances, training labels, all available instances, and side (phenotypic) information, respectively.

Three kinds of subject side information (gender, age, and handiness score) are selected and encoded as the matrix $\mathbf{D}$, where gender is encoded as $1$ (male) and $-1$ (female), and the age and handedness scores are normalized to zero mean and unit variance.
Empirically, Hinge (SVM) loss, $\ell_2$ norm, and Hilbert-Schmidt Independence Criterion (HSIC;\cite{gretton2005measuring}) are employed for $\mathcal{L}(\cdot, \cdot)$, $\|f\|^2_K$, and $\rho(\cdot, \cdot)$, respectively. 
Therefore, the classification algorithm is a semi-supervised SVM trained on the labelled data, with the coefficients regularised by the HSIC between the classifier decision scores and subject phenotypic information.

\subsection*{Submission 4 (S4) - ShefML - Ensemble Model for CNI Challenge}
Team: S. Zhou, M. Kunda, H. Lu

\begin{figure}[ht]
    \centering
    \includegraphics[width=\linewidth]{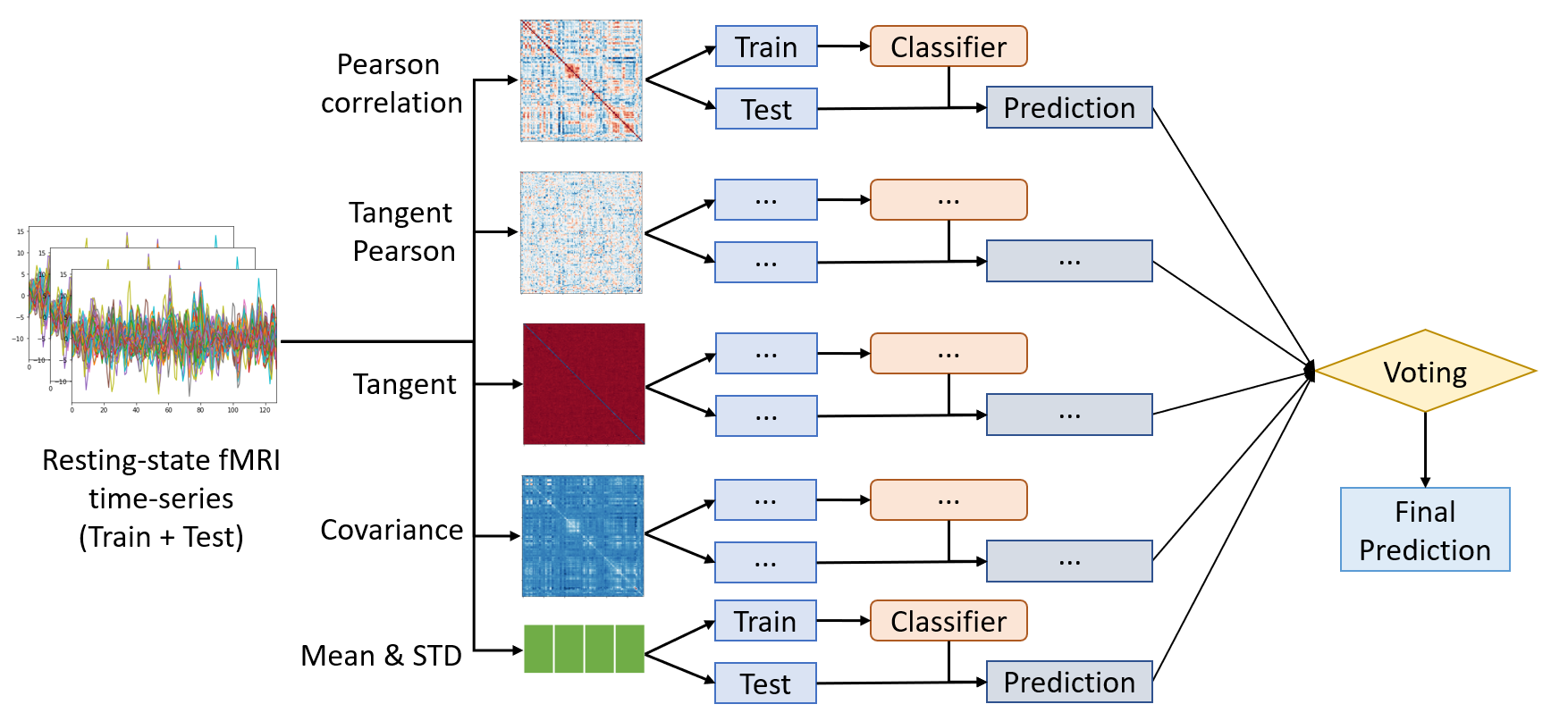}
    \caption{The learning pipeline of Submission 4. Multiple features based on varying definitions of connectome creation were utilized in individual classifiers. Individual predictions were merged using majority voting.}
    \label{fig:s4}
\end{figure}

This approach contains three steps of analysis: \textit{feature extraction, classifier training, and ensemble}. In the first step, five different type of features are extracted via computing the mean and standard deviation, Pearson correlation, tangent (\cite{varoquaux2010detection}), covariance (\cite{varoquaux2010brain}), and tangent Pearson correlation (\cite{kunda2020improving}) from the fMRI time-series data of the AAL atlas. In the second step, five classifiers are trained on each type of feature extracted in step one respectively. The technical details here are exactly the same as in the classification method of S3 (see section~\ref{sec:S3} - Submission 3), which can be viewed as an alternative approach of this solution. In the last ensemble step, the final predictions are made by summarising the predictions given by the five classifiers from step two via majority voting. The pipeline is summarized in Figure \ref{fig:s4}.

\subsection*{Submission 5 (S5) - YaleIPAG - Learning Generalizable Recurrent Neural Networks from Small Task-fMRI Datasets}
Team: N.C. Dvornek, J. Zhuang

\begin{figure}[ht]
\includegraphics[width=1\textwidth]{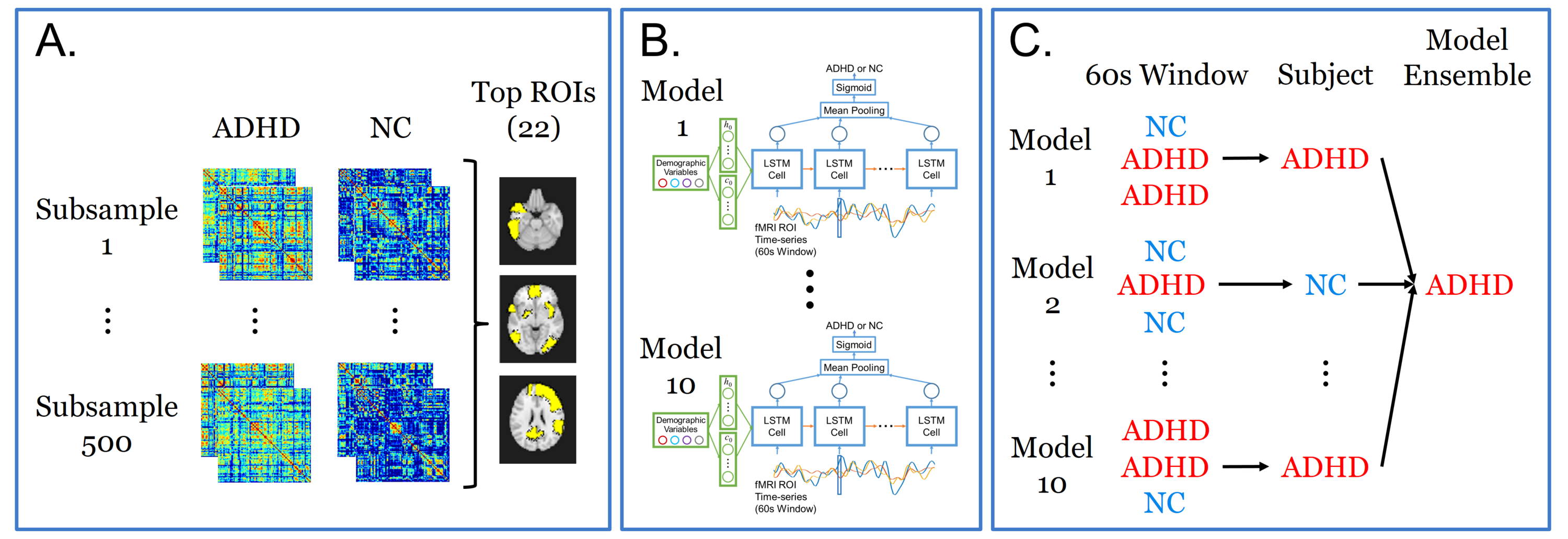}
\caption{\label{fig:pipelineS5}Pipeline overview. A. Top AAL ROIs with pairwise correlations were selected which consistently differed between ADHD/NC using t-tests and repeated subject subsampling (90\%). B. Ten LSTM models were trained using 10-fold split with top ROIs as input. C. Prediction of each subject using all 60s windows and model ensembles.} 
\end{figure}

The method (Fig. \ref{fig:pipelineS5}) is based on \cite{Dvornek2018a}, which aims to learn generalizeable recurrent neural networks from small fMRI datasets. An LSTM-based network learns directly from the ROI time-series data, while the demographic data is incorporated through subject-specific initialization of the LSTM hidden and cell states. 

Using the AAL atlas, ROI selection was first performed by keeping ROIs whose connectivity (i.e., correlation) was consistently significantly different between ADHD and controls groups (two-sample t-test, $p<0.05$) in 500 random subsamples of the data, using 90\% of the training subjects in each subsample. Connectivity was considered consistently different if the $p$-value was in the top 2\% of smallest $p$-values in each of the 500 subsamples. This process resulted in 22 ROIs.

The network architecture consisted of an LSTM layer with $M=32$ hidden units and $T=24$ timesteps (60s window), whose outputs were sent to a shared (across time) fully-connected layer with 1 node, followed by mean pooling and a sigmoid activation function to give the probability of ADHD. Time-series data from the 22 ROIs was used as input to the LSTM (\cite{Dvornek2017}), while demographic data was used for LSTM state initialization (\cite{Dvornek2018a}). Specifically, demographic data was input to two fully-connected layers with $M$ nodes each, representing the initial hidden and cell state of the LSTM.

To improve robustness, 10 models were trained using 10-fold cross-validation splits of the training dataset, with the validation dataset used to determine when to stop training. To predict on a new subject, all possible 60s windows of the time-series were input to the models to get a binary ADHD/control prediction for each window. The subject-level probability of ADHD for a single model was the proportion of windows labeled as ADHD for that model. Finally, the probability of ADHD for a given subject was computed as the mean of the 10 models' predictions.

\section{}
\label{app:C}

An example of pooling the performance from all Teams into an aggregate score by computing the average, median and maximum confidence of metrics across all classifiers. The maximum confidence was calculated by picking the classification where the predicted score is furthest away from 0.5, reflecting the confidence of an algorithm in its classification.

\bgroup
\linespread{1.2}
\begin{table}[ht!]
    \centering
    \caption{Consensus voting results of the challenge summarized as the median of each evaluation metric.  All participants scores were combined using the mean (avg), median, and maximum confidence (maxconf, defined as classification based on the score furthest away from 0.5). In addition, the "Best median metric" out of all submissions from Table~\ref{tab:ranking}) is provided for comparison. The consensus vote does not outperform the best median metric in the ADHD cohort, while performing close to chance in the ASD classification task.}
    \label{tab:appendix1}
    \tiny
    \begin{tabular}{|c||c|c|c|c||c|c|c|c|}
    \hline
        & \multicolumn{4}{c||}{\textbf{ADHD}} & \multicolumn{4}{c|}{\textbf{ASD}} \\ \hline 
        & avg & maxconf & median & Best median metric & avg& maxconf & median & Best median metric \\ \hline
        Acc & 0.35 & 0.48 & 0.35 & \bf 0.66 & 0.50 & \bf 0.55 & 0.50 & \bf 0.55 \\ \hline
        AUC & 0.63 & 0.57 & 0.65 & \bf 0.68 & 0.51 & 0.44 & 0.54 & \bf 0.56 \\ \hline
        F1  & 0.32 & 0.48 & 0.32 & \bf 0.70 & 0.51 & 0.53 & 0.52 & \bf 0.55 \\ \hline
        FDR & 0.67 & 0.52 & 0.67 & \bf 0.33 & 0.50 & \bf 0.46 & 0.50  & \bf 0.47 \\ \hline
        FNR & 0.70 & 0.50 & 0.65 & \bf 0.30 & 0.50 & 0.50 & 0.45 & \bf 0.30 \\ \hline
        FOR & 0.65 & 0.52 & 0.65 & \bf 0.30 & 0.50 & \bf 0.46 & 0.50 & 0.47 \\ \hline
        FPR & 0.65 & 0.50 & 0.65 & \bf 0.35 & 0.50 & \bf 0.45 & 0.55 & \bf 0.45\\ \hline
        GM & 0.33 & 0.48 & 0.33 & \bf 0.70 & 0.51 & 0.53 & 0.53 & \bf 0.56 \\ \hline
        Inf. & -0.30 & -0.05 & -0.30 & \bf 0.35 & 0.00 & \bf 0.10 & 0.00 & 0.05\\ \hline
        Mark & -0.30 & -0.05 & -0.30 & \bf 0.36 & 0.00 & \bf 0.10 & 0.00 & 0.05\\ \hline
        MCC & -0.30 & -0.05 & -0.30 & \bf 0.36 & 0.00 & \bf 0.10 & 0.00 & 0.05\\ \hline
        NPV & 0.35 & 0.48 & 0.35 & \bf 0.70 & 0.50 & \bf 0.54 & 0.50 & 0.53\\ \hline
        OP & 13.86 & 18.95 & 13.86 & \bf 26.96 & 20.00 & \bf 21.82 & 19.90 & 20.86\\ \hline
        Pre & 0.33 & 0.48 & 0.33 & \bf 0.67 & 0.50 & \bf 0.55 & 0.50 & 0.53\\ \hline
        Sen & 0.30 & 0.50 & 0.30 & \bf 0.75 & 0.50 & 0.50 & 0.55 & \bf 0.70\\ \hline
        Spec & 0.35 & 0.50 & 0.35 & \bf 0.65 & 0.50 & \bf 0.55 & 0.45 & \bf 0.55\\ \hline
    \end{tabular}
\end{table}
\egroup

\bibliographystyle{model2-names}\biboptions{authoryear}
\bibliography{references}

\end{document}